\documentclass[aps,pre,preprint,groupedaddress]{revtex4-1}
\usepackage{graphicx}
\usepackage{lineno}
\usepackage{xcolor}

\begin{document}


\title{Bridging discrete and continuous formalisms for biodiversity quantification}


\author{Xue Feng}
\affiliation{ Department of Civil, Environmental, and Geo-Engineering, University of Minnesota, Twin Cities 55455, USA.}
\author{Sara Bonetti}
\affiliation{Department of Environmental Systems Science, ETH Zürich, 8092 Zürich, Switzerland.}
\author{Amilcare Porporato}
\affiliation{Department of Civil and Environmental Engineering and the Princeton Environmental Institute, Princeton, New Jersey 08544, USA.}


\begin{abstract}
Several theoretical frameworks have been proposed to explain observed biodiversity patterns, ranging from the classical niche-based theories, mainly employing a continuous formalism, to neutral theories, based on statistical mechanics of discrete communities. Differences in the descriptions of biodiversity can arise due to the discrete or continuous nature of the underlying models and the way internal or external perturbations appear in their formulations.
Here, we trace the effects of stochastic population dynamics on biodiversity, from the scale of the individuals to the community and based on both discrete and continuous representations of the system, by consistently using measures of community diversity like the species abundance distribution and the rank abundance curve and applying them to both discrete and continuous populations. A novel measure, the community abundance distribution, is introduced to facilitate the comparison across different levels of description, from microscopic to macroscopic. Using a simple birth and death process and an interacting population model, we highlight discrepancies in their discrete and continuous distributions and discuss relevant implications for the analysis of rare species and extinction dynamics. Quantitative consideration of these issues is useful for better understanding of the contributions of non-neutral processes and the mathematical approximations to various measures of biodiversity.
\end{abstract}


\maketitle

\section{Introduction}
Understanding the determinants of species assembly and biodiversity is a central theme of theoretical ecology \citep{Chesson2000,Chave2002,Silvertown2004}; its complexity represents a formidable challenge for modern statistical mechanics \citep{Phillips2006,Goldenfeld2011,Azaele2016}.
The various existing frameworks that seek to explain the origin of observed biodiversity patterns can be classified into two broad categories -- that of classical niche theory and the more recently updated neutral theory, which has received intense attention since being revisited by Bell, Hubbell, and others more than a decade ago \citep{Bell2000,Hubbell2001,Volkov2003}.  
While classical studies on community assembly focus on species-level interactions and differences in their environmental niches \citep{Tilman1982,Chesson2000}, neutral theory adopts a null model in which species are taken to be indistinguishable from each other under all environmental conditions, and community assembly is dominated instead by the probabilistic birth, death, speciation, and immigration occurring at the individual level. Many authors now call for a broader theoretical framework which reconciles neutral and non-neutral processes \citep{Rosindell2012,Matthews2014}, both of which are thought to operate in tandem in natural communities \citep{Chave2004,Leibold2006,Adler2007}. Within this scope, statistical mechanical approaches have been used to explain empirical macro-ecological laws and biodiversity patterns \citep{Dewar2008,McGill2010,Bowler2012,Suweis2012}, paving the way to a reapprochment between neutral and non-neutral frameworks \citep{Dewar2008,Bowler2012}.

One aspect that makes the bridging between neutral and classical population models challenging is the idiosyncrasy of the mathematical formalisms prevailing neutral and niche models. Population dynamics and community assembly have traditionally been analyzed using models with continuous variables that evolve deterministically \citep{May1973,Tilman1982}, while neutral theory has overwhelmingly made use of models for discrete and stochastic variables \citep{Chave2004}. The consolidation of niche and neutral theories necessarily requires extricating descriptions of real processes from their associated modeling tools. This paper aims to clarify the equivalence of continuous and discrete measures of biodiversity. To facilitate the bridging between neutral and niche theories, it also provides novel measures that are easily transferable between discrete to continuous frameworks. The development of statistical tools allowing the comparison of species richness and evenness while accounting for scaling issues and sampling of rare species is key to an accurate description and comparison of population assembly and dynamics \cite{Gotelli2001,McGill2007,Tovo2017}.


The outline of the paper is as follows. Section \ref{sec:links} introduces notation by presenting a brief overview of increasing scales of description for general population models, from detailed microscopic kinetics to deterministic mean behavior. 
Section \ref{sec:CAD} introduces the community abundance distribution, which is then employed in Section \ref{sec:biodmeasures} to derive biodiversity measures from discrete and continuous representations.
In Section \ref{sec:multi_species} multispecies communities from both discrete and continuous population models are constructed and their species abundance distributions and rank abundance curves are compared. Finally, section \ref{sec:discussions} discusses some notable implications for modeling the presence of rare species, mean extinction times, and other aspects of community dynamics and composition.

\section{Background and notation}
\label{sec:links}

Before delving into measures of discrete and continuous populations, it is useful to introduce the underlying population models to ensure consistency of notation throughout the subsequent sections. 
We start at the most fundamental level of individuals, each defined by probabilistic rates of birth and death, and then trace them through progressively higher levels of inference and arriving, in their \textquotedblleft{}thermodynamic limit\textquotedblright{}, at a deterministic, phenomenological description. We compare differences in the transient and long term properties of the solutions derived at each stage of inference, and highlight implications of such differences for quantifying community dynamics and biodiversity. 

A complete characterization of a community at a given time can be generally described by the $S$-dimensional joint probability distribution,
\begin{equation}
p({\bf x})=p_{X_{1},...,X_{S}}(x_{1},...,x_{S}),\label{eq:joint_x}
\end{equation}
such that $X_{1}=x_{1},X_{2}=x_{2},$ and so forth. The variable $X_{k}$ can be considered continuous or discrete. The exact form of equation (\ref{eq:joint_x}) depends on the details of the physical rules defining the growth, death, and interaction between each species in the community (e.g., neutral, mixture, or interacting). Since the high dimensionality of equation (\ref{eq:joint_x}) may render it cumbersome for use, it may be practical to condense the joint distribution into its marginalized distributions, defined for a continuous population as 
\begin{equation}
p_{X_{k}}(x_{k})=\int\int p_{X_{1},...,X_{S}}(x_{1},...,x_{S})\prod_{j\neq k}dx_{j}.\label{eq:marg_x}
\end{equation}
The discrete case is constructed using summations instead of integrals. 

Birth and death processes  \citep{VanKampen1992} provide a natural framework for modeling the
dynamics of populations in systems consisting of discrete units.  
The state of the system is described using the total number of units (e.g., individuals), belonging to one of $S$ fixed groups (e.g., species), is expressed by the multivariate random vector ${\bf X}\equiv X_{1},..,X_{k},...,X_{S}$, whose $k$-th component denotes the number of individuals in species $k$. The probability of $X_{k}$ taking a specific value $x_{k}$ is indicated as $p[X_{k}=x_{k}]$, following notations from \citet{Priestley1981}. A fixed number of $S$ species implies that species evolution is considered to be constrained in such a way that any extinction of a species is exactly balanced by the introduction of a new species.
Changes in the population of each species, $X_{k}$, are caused by an instantaneous jump belonging to one of two processes (birth or death, indicated by $j=1,2$), which are characterized by a transition probability per unit time of moving between two allowable states. The equation that describes the dynamics of the ensemble $p[{\bf X}]$ based on jump probabilities is called the master equation. 

It is often useful to resort to an approximation where the jump processes are replaced with diffusions and the discrete random variable ${\bf X}$ with a continuous random variable (whose probability density we denote with $f({\bf X})$). The rescaling of the physical variables needed to invoke this approximation is analogous to assuming an appropriate ``macroscopic infinitesmal timescale'' (which, incidentally, can be more easily realized for larger populations; see \citet{Gillespie2000} and the rescaling method in \citet{Gardiner2009}). 

At a coarser level of representation, the bulk behavior of the system resulting from averaging the microscopic fluctuations can be described by the macroscopic equation for the mean of ${\bf X}$ as a function of a drift and diffusion term. This can be done in the ``thermodynamic limit'', where system size is taken to infinity while species proportions are kept constant. The result of this procedure, in which the Langevin equation is replaced with yet another set of deterministic ordinary differential equations for ${ \bf X}$ via suitable closure assumptions, is sometimes called a ``phenomenological equation''. It is rarely the case that an exact, closed-form macroscopic equation can be obtained -- this is certainly the case for linear processes with natural boundaries, but not for nonlinear processes; although see e.g., \citet{Ma2015}. The phenomenological equations are thus deterministic population models typified by the Lotka-Volterra type equations in ecology, the SIR models in epidemiology, the Michaelis-Menten reaction kinetics, amongst others (see, for example, \citet{Murray2002}). They represent the bulk behavior of a system, with the embedded assumption that the effect of fluctuations is small relative to the size of the population. It is important to distinguish whether the stochasticity in the original birth and death model originates from ``internal'' or ``external'' sources \citep{VanKampen1992}. The limiting procedure used to derive of the Langevin and Fokker-Planck equations applies only in case of internal noise (e.g., demographic stochasticity \citep{May1973,Lande1993}) through increasing system size. The existence of external noise (or environmental stochasticity) can also change the limiting equations, since in general such extrinsic factors cannot be reduced with increasing system size \citep{Lande1993}.

\section{The Community Abundance Distribution} \label{sec:CAD}
We introduce here the community abundance distribution (CAD) to condense the biodiversity information in the joint community pdf in a reduced form, which can be used to translate the effects of population dynamics into biodiversity measures. The CAD is a bivariate distribution that accounts for both the likelihood
of finding individuals in the community belonging to species $k$ as well as the state of population $x_{k}$ associated with species $k$. It conveniently reduces the number of dimensions from the joint community pdf (from $S$ to two), and captures information in a way that facilitates the calculation of various diversity indices and the species abundance distribution. Because it can be constructed from continuous or discrete populations at any scale of description, it is a very useful starting point for comparing the results of various biodiversity models without being confounded by contrasting assumptions about their population types and driving mechanisms.  

The CAD can be conceptualized through the following: we first build an ensemble of communities by sampling the joint populations from $p_{X_{1},..,X_{S}}(x_{1},x_{2},..,x_{S})$,
in which each sampled point ${\bf x}$ is a realization of a community.
Further sampling an individual from each realization, we consider the bivariate random variable $(K,{\bf X})$ with which it is associated -- the
species index $k$ to which it belongs and the total population of species $x_{k}$ in that community. 
Note that we consider the indices of species, $K$, within the community here as an additional random variable. The CAD is defined as the joint bivariate distribution of the species population ${\bf X}$ and the species index $K$, written as $p_{K,{\bf X}}(k,{\bf x})$. By Bayes' theorem, it is the product of a conditional and a marginal distribution, e.g., 
\begin{equation}
p_{K,{\bf X}}(k,{\bf x})=p_{\mathbf{X|}K}(\mathbf{x|}k)p_{K}(k).\label{eq:joint_xk}
\end{equation} 

This means that (i) the discrete marginal distribution $p_{K}(k)$ of finding species $k$ is equal to the mean proportion of its population in that community, and (ii) the conditional distribution for a given species $k$, $p_{\mathbf{X}|K}(\mathbf{x}|k),$ is equal to the marginal distribution of $X_{k}$ in equation (\ref{eq:marg_x}). These properties
can be summarized as
\begin{equation}
p_{K}(k)=\frac{\langle x_{k}\rangle}{\sum_{k}\langle x_{k}\rangle}, \label{eq:p(k)}
\end{equation}
\begin{equation}
p_{\mathbf{X}|K}(\mathbf{x}|k)=p_{X_{k}}(x_{k}), \label{eq:p(x|k)}
\end{equation}
where $\langle x_{k}\rangle$ is the mean population of species $k$,
given by $\langle x_{k}\rangle=\int up_{X_{k}}(u)du$
for a continuous variable and $\langle x_{k}\rangle=\sum_{u}up_{X_{k}}(u)$
for a discrete variable. 

The resulting form, based on equation (\ref{eq:joint_xk}), becomes
\begin{equation}
p_{K,\mathbf{X}}(k,\mathbf{x})=p_{X_{k}}(x_{k})\frac{\langle x_{k}\rangle}{\sum_{k}\langle x_{k}\rangle}.\label{eq:p(k,x)}
\end{equation}
Thus the CAD concisely captures information from the joint community pdf (equation (\ref{eq:joint_x})) in a way that facilitates the calculation of many diversity measures, most of which are based on the population of different species within the community (encapsulated by eq. (\ref{eq:p(x|k)})) or the proportion of these species (encapsulated by eq. (\ref{eq:p(k)})). 
For example, the marginal distribution  (\ref{eq:p(x|k)}) can used to construct the species abundance distribution, and the species proportions (\ref{eq:p(k)}) for diversity indices. Two popular examples of such indices are the variance-based Simpson's index $\lambda$ and the information theory based Shannon entropy $H$ \citep{Buckland2012,Maurer2011}, expressed as $\lambda =\sum_u p_K^2(u)$ and $H = -\sum_u p_K(u) \ln p_K(u)$. Of course, for the continuous case, summations are replaced by integrals.

\section{Descriptors of biodiversity using continuous and discrete models}\label{sec:biodmeasures}
Much of the recent advances in neutral theory has been buoyed by the ability of species abundance distributions (SADs) to accurately fit relative species abundances. It displays the expected frequency of species at each abundance level, usually binned over logarithmic increments. It can also be found using the CAD by summing over the marginal distributions of all species, e.g.,

\begin{equation}
\langle\phi(x)\rangle=\sum_{k=1}^{S}p_{X_{k}}(x_{k}).\label{eq:SAD}
\end{equation}
This summation is valid for either a continuous or discrete distribution
of $p_{X_{k}}(x_{k})$, since $K$ itself is discrete. When $\mathbf{x}$ is discrete, $\langle\phi(x)\rangle$ represents the mean number of species found with the number of individuals at $x$ \citep{Volkov2003}. If $\mathbf{x}$ is continuous, then the resulting summation represents a ${\it density}$ distribution of size $S$, in which case a more straightforward interpretation is obtained using its cumulative distribution function 
\begin{equation}
\Phi(x)=\int_{0}^{x}\langle\phi(u)\rangle du,
\end{equation}
showing the expected number of species with population at or below $x$. An advantage of using the cumulative distribution $\Phi(x)$ is that it allows for comparison of not only the discrete and continuous solutions, but also of the macroscopic solution for which the population of each species is represented by a Dirac delta function around its mean (which would otherwise be untenable using the density formulation of equation (\ref{eq:SAD})). 

An alternate measure of community biodiversity is the rank abundance curve. When the community composition ${\bf X}$ is a random variable consisting of many realizations, the rank abundance curve can be interpreted as the expected abundance of the $z$-th ranked species over many realizations
\citep{Hubbell2001,MacArthur1960}. Unlike the SAD, the rank abundance curve compares species rank prior to aggregation, and thus requires information directly from the joint community distribution (\ref{eq:joint_x}). It can be derived for continuous populations
as, 
\[
\langle r_{(z)}\rangle=\int\int x_{(z)}p({\bf x})\prod_{k=1}^{S}dx_{k},
\]
where $x_{(z)}$ is the abundance of the $z$-th ranked species
within a realization of $\mathbf{X}$. Discrete populations require
summations in place of integrals. 

For a neutral community, the pdf and the moments of the $z$-th order statistic $X_{(z)}$ can be derived analytically for both discrete and continuous populations using standard tools from order statistics \citep{David2003}. For example, the mean rank abundance (first moment) for a neutral community with discrete populations is given by 
\begin{equation}
\langle r_{(z)}\rangle_{disc}=\sum_{x=1}^{\infty}1-I_{P}\left(z,S-z+1\right),
\label{eq:rank_disc}
\end{equation}
where $I_{P}(\cdot,\cdot)$ is the regularized incomplete beta function of order $P$ \citep{Abramowitz1964} and $P$ is the cumulative distribution function of each discrete marginalized species distribution. Its continuous counterpart is given by 
\begin{equation}
\langle r_{(z)}\rangle_{cont}=\frac{S!}{(z-1)!(S-z)!}\int_0^1F^{-1}(u)u^{z-1}(1-u)^{S-z}du,
\label{eq:rank_cont}
\end{equation}
where $F^{-1}$ is the inverse of the cumulative distribution function of each continuous marginal distribution \citep{David2003}. 

\section{Application to multispecies communities}
\label{sec:multi_species}
We now construct multispecies communities from discrete and continuous population models derived at different scales of representation and compare their associated measures of biodiversity. These communities are evaluated in the context of their biodiversity measures, which are derived using both continuous and discrete formalisms. 

Each community is broadly designated into one of three groups -- neutral, mixture, or interacting -- based on the relationship between its constituent species, though real communities exhibit overlapping features from one or more of these groups. In the neutral community, the population of an isolated species can be extrapolated to account for that of every species in the community. The mixture community is built from the superposition of species which do not interact with other species, but maintain different birth and death rates; in other words, their niches are defined by fundamentally non-interactive, environmentally-based variables \citep{Soberon2007}. The interacting community is characterized by biotic interaction, and is exemplified here by a  competitive Lotka-Volterra model. 

\subsection{Neutral communities}

The neutral community consists of many individuals who are governed by identical probabilistic rates of birth, death, immigration, and speciation \citep{Adler2007,Azaele2016}. A neutral community can be constructed from any one-dimensional (i.e., one species) population model by replicating its result identically for each species in the community. We do this in both the discrete and continuous case, which we then compare in terms of biodiversity measures.

For the discrete case, we consider the special case of density dependent growth and death rates,  i.e., $b(x)=b_{1}x$ and $d(x)=d_{1}x$, which are valid for $x>0$, with $b_{1}, d_{1}$ assumed as constants. To maintain a finite population and to prevent total extinction when $b_{1}<d_{1}$, the lower boundary at $x=1$ is modified to a reflecting boundary, which allows an artificial process to inject new individuals at $x=1$ when the original process reaches $x=0$, ensuring that a new individual can arise from extinction. 
We focus on the steady state solution to the master equation under a reflecting boundary, which can be found using detailed balance as a logarithmic (or log-series) distribution, 
\begin{equation}
p^{st}[x]=\mathcal{N}_{p}\frac{(b_{1}/d_{1})^{x}}{x},\label{eq:sol_log}
\end{equation}
where the normalizing constant is $\mathcal{N}_{p}=-1/\ln\left(1-\frac{b_{1}}{d_{1}}\right)$ for $x=1,2,...,\infty$. 
For the continuous case, the solution to the associated Fokker-Planck equation with reflecting boundary at $x=1$ is
\begin{equation}
f^{st}(x)=\mathcal{N}_{f}\frac{e^{2\frac{b_{1}-d_{1}}{b_{1}+d_{1}}x}}{x}.\label{eq:sol_exp}
\end{equation}
This is an approximation to equation (\ref{eq:sol_log}) for $x\geq1$
with $\mathcal{N}_{f}=1/\Gamma\left(0,2\frac{d_{1}-b_{1}}{d_{1}+b_{1}}\right)$, where $\Gamma(\cdot,\cdot)$ is the incomplete gamma function \citep{Abramowitz1964}. 

\begin{figure}
\includegraphics[width=135mm]{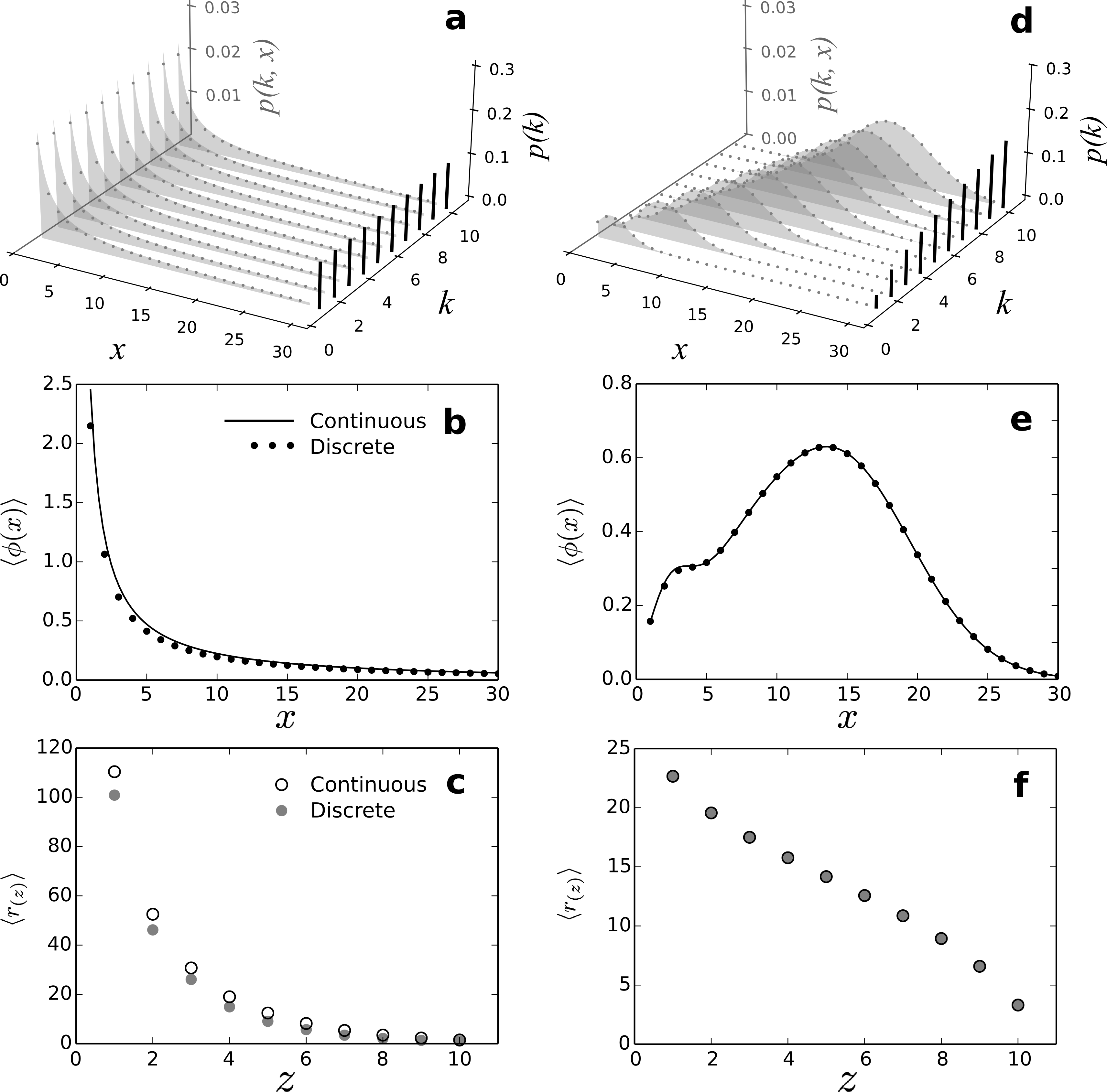}
\caption{The community abundance distribution and biodiversity plots are compared between a neutral community (a,b,c) and a mixture community (d,e,f), for both continuous and discrete populations. The marginal distributions for each species are stationary distributions of a linear birth and death process, with $b_1=0.99$ and $d_1=1.0$ for each species in the neutral community, and $b_1$ values equally spaced between $1.0$ and $40.0$ and $d_1=\sqrt{b_1}/3$ for $S=10$ species in the mixture community. In the community abundance distribution of panels a and d, the discrete population based on equation (\ref{eq:sol_log}) and the continuous population based on (\ref{eq:sol_exp}) are shown respectively with dots and filled polygons, and the proportion of each species in the community, $p(k)$, is projected onto a plane. Panels b and c show respectively the species abundance distribution (equation (\ref{eq:SAD})) and the rank abundance distributions (equations (\ref{eq:rank_disc}) and (\ref{eq:rank_cont})) for the neutral community. Panels e and f show the same for the mixture community. \label{fig:neutralmixture}}
\end{figure}

An example of the resulting neutral community built from the one-dimensional birth and death process  (equation (\ref{eq:sol_log}) and (\ref{eq:sol_exp})) is shown in the CAD of Figure \ref{fig:neutralmixture}a. This is used by Hubbell's neutral model to construct species abundance distributions related to the metacommunity \citep{Hubbell2001,Volkov2003}. The artificial reflecting boundary enacted for this birth and death process is physically associated with a speciation process based on ``point mutation,'' in which new species arise from absolute rarity (as opposed to ``random fission'', in which new species arise from preexisting species of any abundance). It is worth noting, however, that such a mode of speciation makes sense only in the context of an aggregation of ${\it unlabeled}$ species (as adopted by Hubbell). What this process represents for a labeled species is unclear, since it induces a pathological ``resurrection''
only after a species has become extinct at $x=0$ (see however \citet{Vallade2003} and \citet{Chisholm2014} for alternative labels in terms of speciation time).  

The species abundance distribution and the rank abundance curve of neutral community governed by linear birth and death rates are compared in Figure \ref{fig:neutralmixture}b and \ref{fig:neutralmixture}c for the discrete and continuous cases.
The continuous species abundance distribution predicts higher mean species $\langle \phi(x)\rangle$ for smaller $x$ compared to its discrete analogue. Likewise, the rank abundance curve derived from continuous populations also overestimates the expected abundance $\langle r_{(z)}\rangle$ of the highest ranked species (with the lowest $z$ values). These differences in the biodiversity measures are related to the fact that, even though the community size as a whole may be large, the population of each species remains small; a community composed of many rare species. 

\subsection{Mixture communities}

A mixture community can be assembled as a superposition of species that remain non-interacting. Its ecological interpretation is grounded in the concept of Grinnellian niches \citep{Grinnell1917}, described by a class of non-interactive variables pertaining to environmental or historical conditions, and are relevant to understanding the broad scale ecological properties and the geographical extent of each species. These variables, including mean temperature and precipitation, potential evapotranspiration, solar radiation, latitude, and topography, can predict patterns of species richness at the global scale \citep{Francis2003,Kreft2007,Bonetti2017}. Grinnellian niches are contrasted to the class of Eltonian niches \citep{Elton1927} that emphasizes biotic interactions like competition, predation, and mutualism, and require local scale resolution of resource-consumer mechanisms. While both types of niche variables are thought to interact in natural communities, the successful use of Grinnellian niches to predict species distributions at large spatial scales suggests that the distinction between them can be valid and useful \citep{Soberon2007}. 

Here, the mixture community is built from species that can be distinguished solely through their Grinnellian niches. Mathematically, this means that the population of each species in the mixture community can still be modeled using one-dimensional birth and death processes, only now with potentially different birth or death parameters for each species. Thus, in the mixture community, each species has varying rates of birth and death. Figure \ref{fig:neutralmixture}d illustrates an example of a mixture community built from species whose population is governed by a linear birth and death process, in which growth is now determined solely by immigration and decay is through density dependent deaths, i.e., $b(x)=b_{0}$ and $d(x)=d_{1}x$ where $b_{0},d_{1}>0$. In this case, the solution to the master's equation for the linear birth and death process becomes a Poisson distribution, 
\begin{equation}
p^{st}[x]=\mathcal{N}_{p}\frac{(b_{0}/d_{1})^{x}}{x!},\label{eq:sol_poisson}
\end{equation}
defined for $x=0,1,...,\infty$, with $\mathcal{N}_{p}=e^{-b_{0}/d_{1}}$. This equation can be approximated through the solution to the Fokker-Planck equation as 
\begin{equation}
f^{st}(x)=\mathcal{N}_{f}e^{-2x}\left(\frac{d_{1}}{b_{0}}x+1\right)^{4\frac{b_{0}}{d_{1}}-1},\label{eq:sol_gamma}
\end{equation}
for continuous realizations of $x\geq0$, with $\mathcal{N}_{f}=\left(d_{1}/b_{0}\right)e^{-2b_{0}/d_{1}}/E_{1-4\frac{b_{0}}{d_{1}}}(2b_{0}/d_{1})$
where $E_{n}(\cdot)$ gives the exponential integral function of order $n$ \citep{Abramowitz1964}. 
The species abundance distribution and the rank abundance curves for this mixture community (Figure \ref{fig:neutralmixture}e and \ref{fig:neutralmixture}f) are well matched between discrete and continuous formulations, since the likelihood of finding a rare specie in this particular community is small. 

\begin{figure}
\includegraphics[width=135mm]{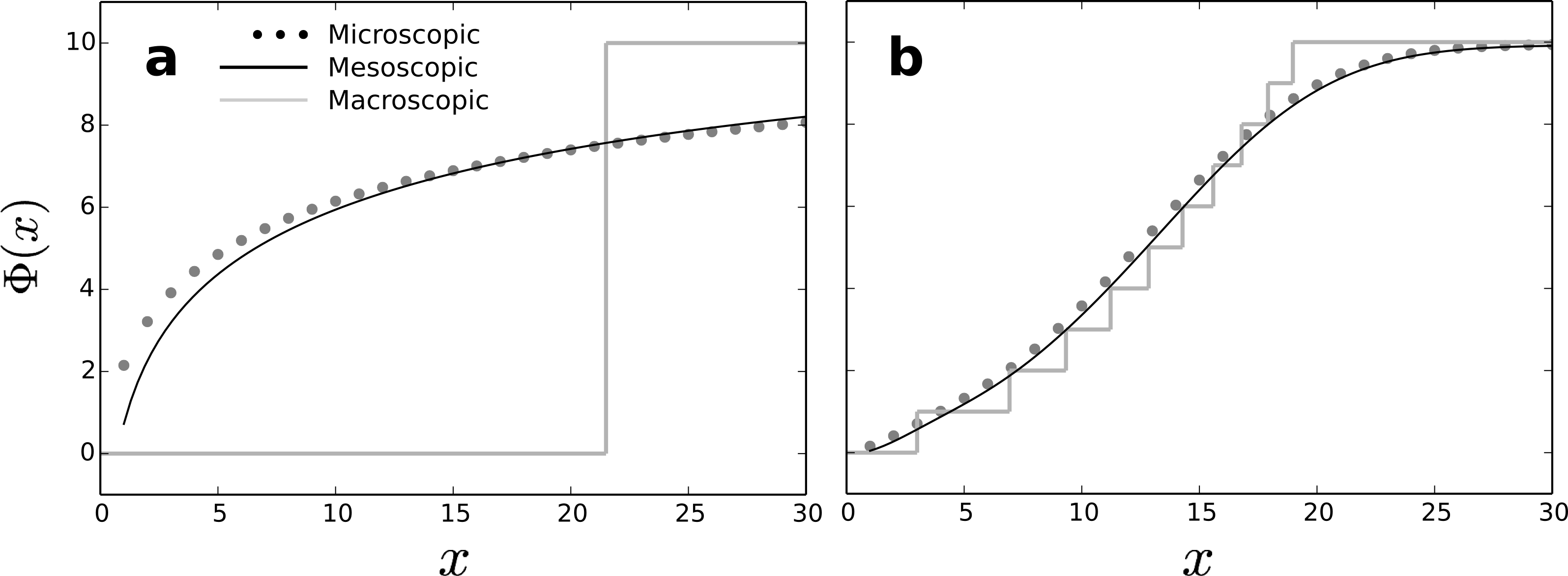}
\caption{The cumulative species abundance, $\Phi(x)$, is shown for the neutral (a) and the mixture (b) community of Figure \ref{fig:neutralmixture}. The discrete and continuous solutions are derived respectively from the master equations (\ref{eq:sol_log}) and (\ref{eq:sol_poisson}), and the Fokker-Planck approximations (\ref{eq:sol_exp}) and (\ref{eq:sol_gamma}). The deterministic solutions are constructed based on macroscopic, determinist species population. Each population is represented by a Dirac delta function centered on the mean of master equation. Thus, the discrete increments in $\Phi(x)$ for the deterministic solutions result not from population variabilities but from the occurrence of a species with a distinct Grinnellian niche. \label{fig:cumSAD}}
\end{figure}

The descriptions of the full community at its microscopic, mesoscopic, and macroscopic scales can be compared for discrete and continuous versions using the cumulative abundance distribution $\Phi(x)$ (Figure \ref{fig:cumSAD}). Previously, the continuous and discrete formulations have been used to respectively describe populations at the mesoscopic and microscopic scale. Now, the pdf of a population described macroscopically (i.e., by its mean) is represented by a Dirac delta function centered around its mean; its cumulative distribution function (cdf) is a Heaviside step function. Because the macroscopic description allows for no variability at the population level, changes in the corresponding macroscopic cumulative abundance distribution $\Phi(x)$ must be explained by differences in the species' Grinnellian niches. For example, in a neutral community, the macroscopic $\Phi(x)$ consists of a single step because all species effectively occupy the same niche (Figure \ref{fig:cumSAD}a) whereas for a mixture community, the total abundance of the community is made up incrementally by species differing in their Grinnellian niches (Figure \ref{fig:cumSAD}b). Thus, the relative contributions to the overall community abundance from variability in either population (e.g., through $X_k$) or species level (e.g., through $K$) can be examined by plotting the cumulative abundance distributions, enabling the comparison of biodiversity measures across different levels of description.

\subsection{Interacting communities} \label{LV}

In comparison to neutral and mixture communities, even the most simple communities containing Eltonian niches \citep{Elton1927}, which call for species interactions, will result mathematically in nonlinearities and couplings in its microscopic dynamics which render its complete solution analytically intractable. We use here a ten species ($S =10$) Lotka-Volterra model and trace its evolution from the reaction kinetic formulation to its macroscopic description based on a set of well-known ordinary differential equations \cite{May1973,Qian2011}. Its discrete and continuous solutions are simulated and compared. 

Let  ${\bf X_i(t)} = [X_1(t),...,X_i(t),...,  X_S(t)]$ be the populations for species $i=1,..,S$, $r_i$ their intrinsic rates of growth, $\alpha_{ii}$ are parameters corresponding to strengths of per capita intraspecific competition, and $\alpha_{ij}$ for interspecific competition. 
Following the large population limit, we can cast the Lotka-Volterra model as a set of Langevin equations \citep{Gardiner2009}, i.e.,  
\begin{eqnarray}
\frac{dX_i(t)}{dt}=A_i(\mathbf{X})+B_i(\mathbf{X})W
\label{eq:lang_LV}
\end{eqnarray}
where $A_i(\mathbf{X}) = r_i X_i-\alpha_{ii}X_i^2-\alpha_{ij}X_i X_j$ and $B_i(\mathbf{X})=\sqrt{r_i X_i+\alpha_{ii}X_i^2+\alpha_{ij}X_i X_j}$ are the drift and diffusion term, respectively.
Equations (\ref{eq:lang_LV}) are often proposed as a stochastic counterpart, under natural boundary conditions, of a deterministic set of ordinary differentials equations describing the competition between $S$ species (e.g, \citet{ May1973,Qian2011}), 
\begin{eqnarray}
\frac{dx_i(t)}{dt}=r_i x_i(t)-x_i(t) \sum_{j=1}^S \alpha_{ij}x_j(t).
\label{eq:phenom_LV}
\end{eqnarray}

This set of phenomenological equations describes the dynamics of the $S$ species at their thermodynamic limit. By comparing them against equations (\ref{eq:lang_LV}), it becomes clear that their equivalence can only be established by setting $x=\langle X\rangle$, $y=\langle Y\rangle$, $x^2=\langle X^2\rangle$, $y^2=\langle Y^2\rangle$, and $xy=\langle XY\rangle$. As a result the long term behavior of the deterministic equations (\ref{eq:phenom_LV}) and the Langevin equations can differ substantially.

\begin{figure}
\includegraphics[width=135mm]{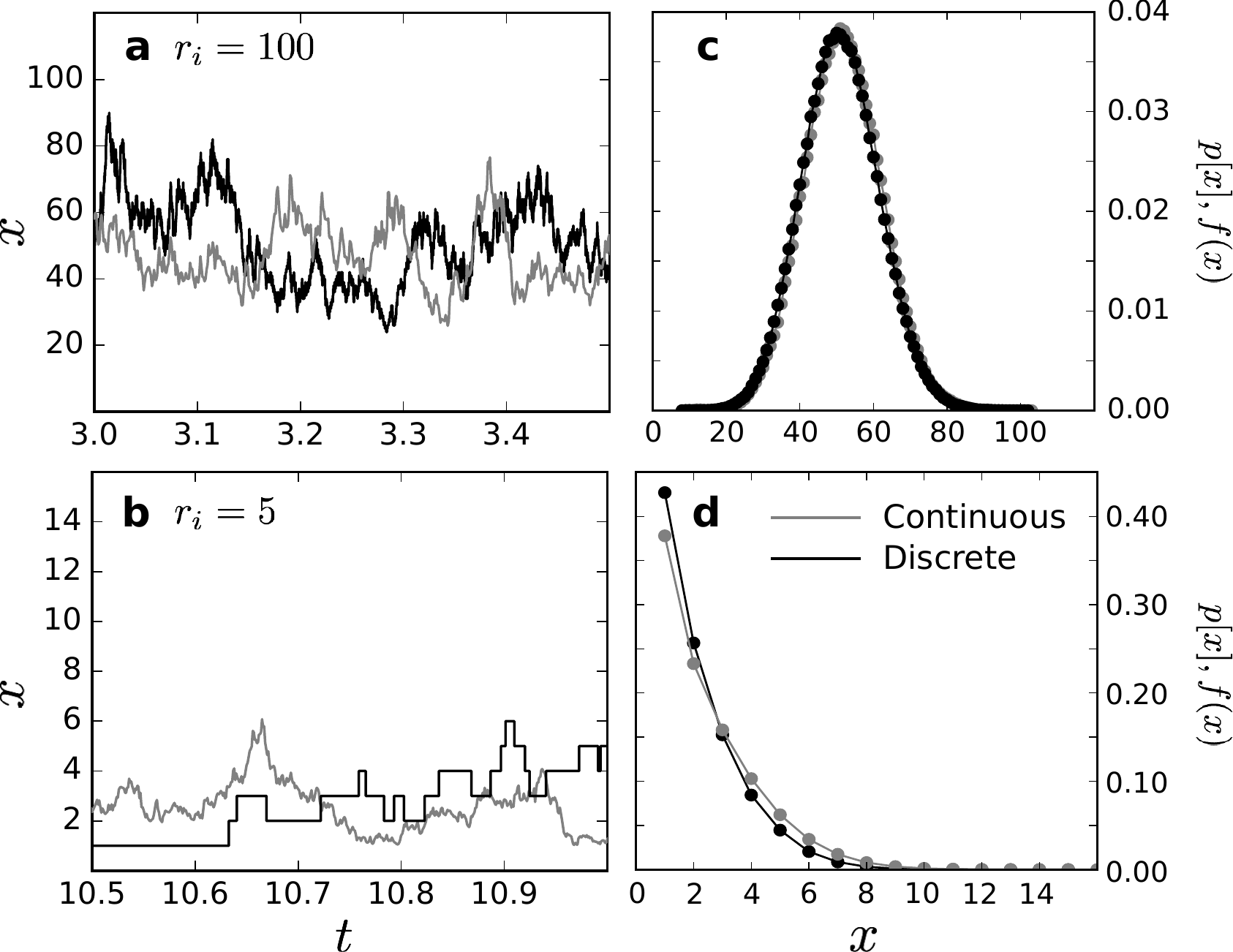}
\caption{Simulated trajectories (a,b) and binned histograms (c,d) are shown on based the ten-species competitive Lotka-Volterra model (Section \ref{LV}).  The discrete and continuous simulations are based respectively on the stochastic kinetic reactions (black lines) and the Langevin equations (\ref{eq:lang_LV}) (grey lines). Top panels (a,c) show a process with $r_1=100$ and $\alpha_{ij}=0.1$ and bottom panels (c,f) are for $r_1=5$ and $\alpha_{ij}=0.1$. Both processes use $\alpha_{ii}=1.0$. Results are shown only for one of the ten species.  \label{fig:sim_LV}}
\end{figure}

The stationary distributions from the master equation and the Langevin equation can be compared by erecting an artificial reflecting boundary for the system at $x=1$ (Figure \ref{fig:sim_LV}). As expected, their trajectories are qualitative similar at larger populations, and the Langevin solution is a good approximation for the stationary distribution. The solutions diverge substantially at smaller populations, with a higher probability of small values predicted by the continuous approximation. 

\begin{figure}
\includegraphics[width=135mm]{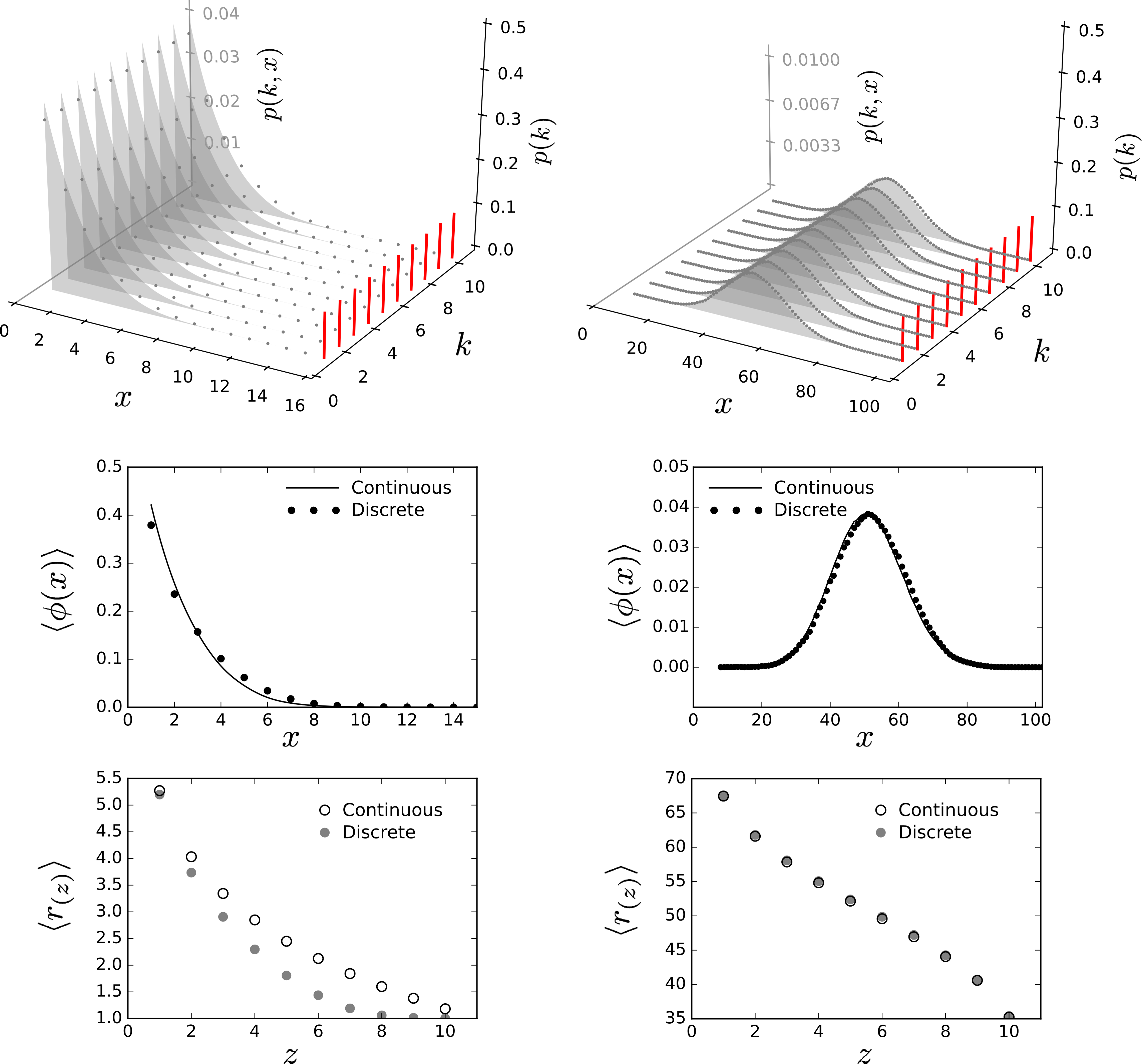}
\caption{The community abundance distribution and biodiversity plots for interacting communities for both continuous and discrete populations. Left panels show results with $r_i=5$, $\alpha_{ii}=1$, $\alpha_{ij}=0.1$, and $x_{i}(0)=10$, right panels are for $r_i=100$, $\alpha_{ii}=1$, $\alpha_{ij}=0.1$, $x_{i}(0)=50$. 
In the community abundance distribution (top panels), the discrete and continuous populations are shown respectively with dots and filled polygons, and the proportion of each species in the community, $p(k)$, is projected onto a plane. Middle and bottom panels show respectively the species abundance distribution and the rank abundance distributions.}
\label{fig:LVcad}
\end{figure}

\begin{figure}
\includegraphics[width=135mm]{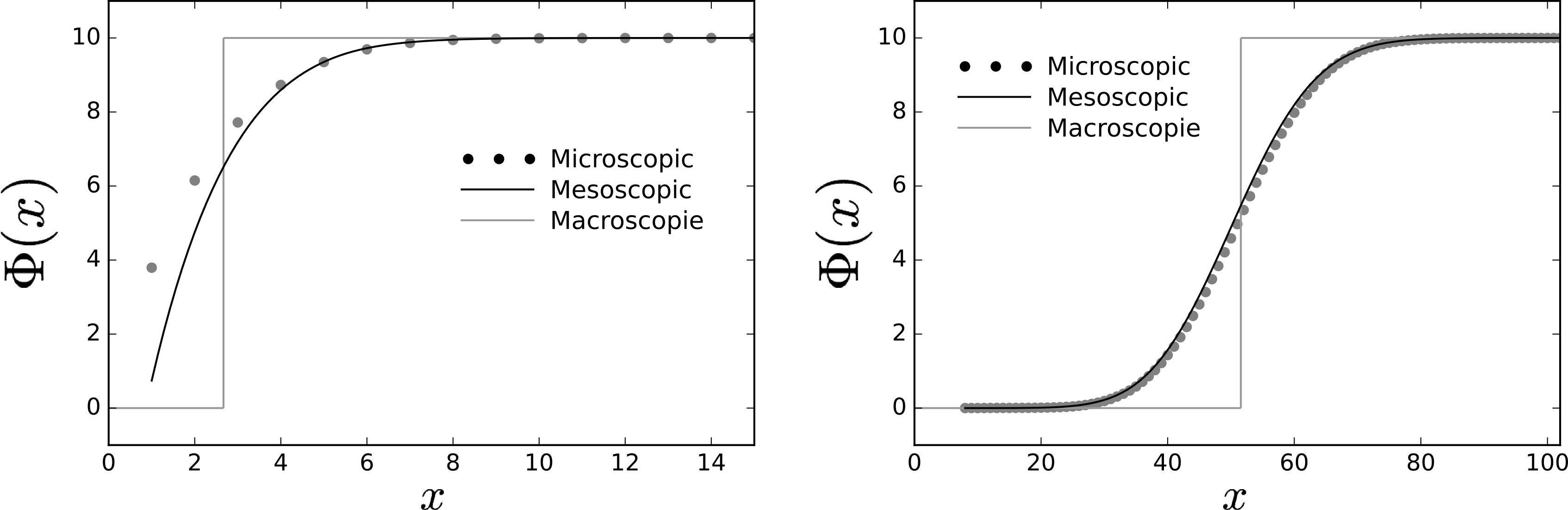}
\caption{The cumulative species abundance, $\Phi(x)$, is shown for the interacting communities of Figure \ref{fig:LVcad} (left panel is for $r_i=5$, $\alpha_{ii}=1$, $\alpha_{ij}=0.1$, and $x_{i}(0)=10$, right panel for $r_i=100$, $\alpha_{ii}=1$, $\alpha_{ij}=0.1$, $x_{i}(0)=50$). The deterministic solutions are constructed based on macroscopic, determinist species population.}
\label{fig:LVcum}
\end{figure}

The resulting communities are shown in Figures \ref{fig:LVcad}, together with their species abundance distributions and rank abundance curves, while Figure \ref{fig:LVcum} compares the cumulative abundance distributions for microscopic, mesoscopic, and macroscopic descriptions. 
Both the species abundance distribution and the rank abundance curves predict higher number of lower population (i.e., rare) species in the case of the continuous description, while good agreement is attained between the continuous and discrete framework at larger populations. Again, the macroscopic cumulative abundance distribution $\Phi (x)$ consists of a single step because all species occupy the same niche.

\section{Discussion and Conclusions}\label{sec:discussions}

We compared the cumulative species abundance $\Phi(x)$ of microscopic, mesoscopic, and macroscopic descriptions simultaneously in Figures \ref{fig:cumSAD} and \ref{fig:LVcum}, illustrating the differences resulting both from using a large population approximation and from assuming negligent variability in the population of each species within the community. 
What this means in terms of modeling biodiversity is that the results will reflect model preferences for the scale of interest, though this fact is rarely acknowledged explicitly. 
From a practical perspective, the diversity of a community quantified by sampling from true populations can be interpreted either as population means or single realizations, thus influencing how the biodiversity of that community will be represented. By introducing the CAD, we aim to facilitate the comparison of measured diversity of different communities constructed from various interpretations.

As an example, the number of rare species may be significantly misrepresented when using a continuous approximation. We have shown that the Fokker-Planck approximation markedly overestimates the likelihood of small population sizes, as the continuous approximation to the Lotka-Volterra model agrees better with its discrete solutions only at large populations (Figure \ref{fig:sim_LV}). These and similar results have been obtained previously in the classical studies (e.g., \cite{Ethier1977}), where errors due to the diffusion approximation in the Fokker-Planck equations where shown to be inversely proportional to the population size. We emphasize here, however, that these differences can be propagated to community measures derived from these species populations and can persist even if the total community size is large, as long as each individual species abundance is low. These differences are consequences of modeling assumptions, and should not be conflated with the many physical explanations for why more rare species are observed in large assemblages when compared to predictions -- such as via the distinction between core and occasional species \citep{Magurran2003} -- as well as statistical explanations based on sampling methodologies \citep{McGill2003}.

The discrepancies between continuous and discrete formulations can be inspected not only through their stationary distributions but also through their transient dynamics. At smaller populations, the discrete jumps are larger and occur less frequently compared to their diffusion counterparts, thus distinguishing them as qualitatively different processes (Figure \ref{fig:sim_LV}). The relative sizes of these discrete jumps are masked only at larger populations. As a result, the use of the large population approximation can confound the length of the mean extinction time, which by definition involves excursions close to the extinction threshold. Since the Fokker-Planck approximation in general fails to correctly account for large fluctuations that result in extinctions \citep{Ovaskainen2010}, to describe extinction time, species age, and extinction likelihood, it may be necessary to work directly from the master equation \citep{Chisholm2014} or adopt a different approximation method like the Wentzel-Kramers-Brillouin (WKB) approximation \citep{Bender1978}, which is more suited to rare event statistics. 

The measures of biodiversity introduced here can be adopted for both continuous and discrete population models, allowing an objective comparison between niche and neutral frameworks. 
In particular, both discrete and continuous representations of the system were linked to measures of the community diversity  with limiting solutions of both neutral and non-neutral (mixed or interacting) communities to emphasize the separation of process from observed patterns. Ultimately, we hope that these considerations will contribute to providing a stronger theoretical foundation for understanding the role of neutrality and quantifying the effects of niches on various measures of biodiversity.

\begin{acknowledgments}
SB and AP acknowledge support from US National Science Foundation (FESD EAR-1338694). AP also acknowledges support from the USDA Agricultural Research Service cooperative agreement 58-6408-3-027; and National Science Foundation (NSF) grants CBET-1033467, EAR-1331846, EAR-1316258, and the Duke WISeNet Grant DGE-1068871. XF is supported by the NOAA Climate and Global Change Postdoctoral Fellowship Program. 
\end{acknowledgments}

\bibliography{BiodMeasuresBiblio}

\end{document}